\pdfoutput=1
\documentclass[10pt]{article}

\usepackage{algorithm}
\usepackage{color}
\usepackage{bbm}
\usepackage{epsfig}
\usepackage{amssymb,amsmath,amscd, mdwmath}
\usepackage{graphicx,cite,url}
\usepackage{algorithmic}
\usepackage{float}
\usepackage{amsfonts}
\usepackage{syntonly}
\usepackage{multirow}
\usepackage{graphicx}
\usepackage{subfig}
\usepackage{tabularx}
\usepackage{setspace}
\usepackage{stfloats}
\usepackage{amsthm}
\usepackage[labelfont=bf,labelsep=period,justification=raggedright]{caption}

\makeatletter
\renewcommand{\@biblabel}[1]{\quad#1.}
\makeatother

\date{}

\begin{document}

\begin{flushleft}
{\Large
\textbf{Predicting publication productivity for researchers: a piecewise Poisson model
 }
}
\\
Zheng Xie$^{1,2, \sharp }$
\\
\bf{1}   College of Liberal Arts and Sciences,   National University of Defense Technology, Changsha,   China. \\
\bf{2} Department of
Mathematics, University of California, Los Angeles,   USA
\\  $^\sharp$ xiezheng81@nudt.edu.cn
 \end{flushleft}
\section*{Abstract}
Predicting the scientific productivity of researchers is a basic task for academic administrators and funding agencies. This study provided a model for the publication dynamics of researchers, inspired by the distribution feature of researchers' publications in quantity. It is a piecewise Poisson model, analyzing and predicting the publication productivity  of researchers by regression. The principle of the model is built on the explanation for the   distribution feature as a result of an inhomogeneous Poisson process that can be approximated    as a piecewise
Poisson process.
 The model's principle was validated by the high quality dblp dataset,   and  its effectiveness was testified  in predicting the publication productivity for majority of researchers and  the evolutionary trend of their publication productivity.
 Tests to confirm or disconfirm the model are also proposed. The model has the advantage of providing results in an unbiased way; thus is useful for funding agencies that evaluate a vast number  of applications with a quantitative index on publications.




%


\noindent {\bf Keywords:}  Scientific publication,  Productivity prediction, Data modelling.

\section*{Introduction}


%

Scientific fields such as informetrics, scientometrics, and bibliometrics establish a range of  models and methods  to evaluate the impacts of scientific publications, and then to predict the scientific success of  researchers\cite{Sinatra2016}.
Although publication productivity     correlates to
scientific success, much attention on this topic has concentrated  on  citation-based indexes,   followed by
     the $h$-index provided by Hirsch\cite{Hirsch2005}, a  popular  measure of  scientific success.
It  is
       the maximum value of $h$ such that a researcher   has produced $h$ publications that have each been cited at least $h$ times.
The  popularity of the $h$-index is attributable  to its simplicity and its addressing both the productivity and the citation impact of  publications\cite{Schubert2007}.

  The success in the prediction of  citation-based indexes  and the $h$-index  can be thought to result from  the  cumulative advantage of receiving  citations found by Price in the year 1965\cite{Price1965}, which   has been extended  as a general theory for bibliometric and other cumulative advantage process\cite{Price1,Barabasi1999a,PercM2014}. From the perspective of   statistics,  the success is due to  the  predictable components  of these indexes    that can be extracted via autoregression. In more concrete terms,
   the current $h$-index,
  the number of annual
citations,   and   the number of    five-year citations
  are  found to be   positive predictors to these  indexes\cite{Acuna2012,Newman2014}.

  The cumulative advantage in producing publication    is weaker than that in receiving citations.
Empirical studies on data display   it as the phenomenon   that
 the tail  of the quantitative distribution of the publications produced by a group of researchers   is much shorter than that of the citation distribution  of those researchers\cite{XieX2017}.
This study also shows   the productivity of researchers in the dblp dataset
does not have a  predictable component  that can be extracted via autoregression.
 The  autocorrelation coefficients with a lag larger than $1$ of the time series   on an individual  researcher's cumulative publication quantity   are almost smaller than $0.5$,   suggesting a   lack of predictability in a researcher's productivity only by his or her historical  publication quantity.
 Therefore, the critical factor of the   success in the prediction   of citation-based indexes and the $h$-index
does not exist   in the prediction task of publication productivity.





 Is there any  predictability in  publication patterns?
 Given the  factors involved in publishing, such as the intrinsic value of research work, timing,   and the publishing venue, finding regularities in the publication history of   researchers is an elusive task. Age and achievement probably constitute the most comprehensive   attempt to empirically determine the   changes in   researchers' creativity,
reflected by the changes in
their publication productivity\cite{Lehman2017}.
In network science, these factors are called the aging phenomenon and the cumulative advantage,  dominating the evolution of coauthorship networks\cite{Glanzel2,XieO2018}. Hence,
the productivity has been theoretically expressed as a curvilinear function of age\cite{Simonton1984}. This theoretical result  is   suitable for the fruitful  researchers that have a long time
engaging in research.
However,  it
  cannot fit   the productivity evolution of   many researchers  in the dblp dataset analyzed here.


Empirical datasets from several disciplines show that the number of a researcher' publications approximately follows a generalized Poisson distribution with a fat tail\cite{XieLL2018}. Can this feature be reproducible by dynamical random models? Previous studies show this distribution  can be  thought as      a mixture of   Poisson distributions\cite{XieO2016}. Samples following the same  Poisson distribution means that they would be   drawn from the same population. It means researchers can be partitioned into several populations, such that
 each population has certain homogeneity  in publication patterns. Finding such a  partition would help  to reveal the mystery of publication patterns, which inspires this study.


We
partitioned the researchers in the dblp dataset into several subsets, 
each of which
 consists of the researchers with the same number
of historical publications produced before the current time. 
 This partition eliminates the diversity in publishing experience. For each subset, we found that the number of publications of a member
follows a Poisson distribution
  at   the following short time interval.   This  inspires us to provide a piecewise Poisson  model to find significant predictors for publication  productivity.
   The finding is that   researchers' publication  productivity
    significantly correlates to time given a  historical publication quantity.
  The relationship allows us to infer researchers' publication  productivity in the future. We provided two methods to test the prediction results of our  model  in terms of    the evolutionary trend of researchers' productivity and the quantitative distribution of their  publications.


This paper is organized as follows.  Literature review and empirical data are described in Sections 2, 3. The model is described in Sections 4-6, where the experiments   and   comparisons with previous results   are also analyzed. The results are discussed and conclusions drawn in Section 7.



\section*{Literature review}

There
are three main  aspects  to the prediction of  scientific success: the $h$-index,     citation-based indexes, and    publication productivity. Although our study focuses on predicting the publication  productivity of researchers,  reviewing the methods in first two aspects contributes to finding the possibility and unavailability of applying those methods to the third aspect.

 As a popular measure of scientific success,  the $h$-index of researchers attracts  considerable attention on predicting it. Acuna et al analyzed
the data of 3,085 neuroscientists,
57 Drosophila researchers, and 151 evolutionary scientists by   a linear
regression with elastic
net regularization. They  presented a formula to predict the
$h$-index,  and indicated that the current $h$-index is the most significant  predictor, compared
with the number of current papers, the year  since publishing first paper, etc\cite{Acuna2012}.
  Dong
et al   utilized   the  standard linear  regression and logistic regression to analyze more  features, such as  the average  citations of an author's papers and the number of coauthors\cite{Dong2016}.
  Mccarty  et al analyzed  the coauthorship  data of  238 authors collected from the Web of Science,  and showed  that the number of
 coauthors and their  $h$-index also are positive  predictors\cite{Mccarty2013}.


The number of received citations is a widely-used measure of success for publications and   researchers. To predict highly cited publications only based on short-term citation data, Mazloumian applied a multi-level regression model\cite{Mazloumian2012},  Wang et al
 derived a mechanistic model\cite{Wang2013},  Newman defined $z$-scores\cite{Newman2014},
 Gao et al   utilized  a Gaussian mixture model  \cite{Cao2016},   Pobiedina applied   link prediction\cite{Pobiedina2016}, and Abrishami et al utilized  deep   learning\cite{Abrishami2019}. Together with the impact factors of journals, Stern and Abramo  applied linear regression models to this prediction task respectively\cite{Stern2014,Abramo2019},  and Kosteas   introduced the rankings of
 journals\cite{Kosteas2018}.
 To improve prediction precision,   the information of authors and contents of publications are utilized:
Bornmann et al added
  publications' length\cite{Bornmann2014};
Bai et al applied
maximum likelihood estimation,  and introduced the aging  of   publications' impact\cite{Bai2019};
   Sarigol et al  used  a  method of random decision forests, and introduced specific  characteristics of coauthorship networks (e.~g.  the  centrality)\cite{Sarigol2014};
 Yu et al provided    a stepwise  regression model   synthesizing specific  features of publications, authors, and   journals\cite{YuYu2014};
 Klimek et al utilized
the centrality measures of
term-document networks\cite{Klimek2016}.

Returning to the prediction of publication productivity, one may find  that the studies  on  this aspect are quite few when compared with those on $h$-index and  citation-based indexes. Empirical studies found the cumulative advantage in producing publications and the aging of researcher' creativity\cite{Newman2001,TomassiniM2007}. Laurance et al analyzed the publications of 182 researchers by using the Pearson correlation coefficient, and found that  Pre-PhD publication success   strongly  correlated to long-term success\cite{Laurance2013}. In the aspect of theoretical research, Lehman concluded that achievement tends to be a curvilinear function of age. From the onset of a researcher's career, productivity tends to rapidly increase, then reaches at the peak productive age, and thereafter slowly declines with  aging\cite{Lehman2017}. Simonton provided a formula to model this process\cite{Simonton1984}.

 The  aforementioned methods  of citation-based indexes and $h$-index  all refer to the positive   correlation between the current indexes and their history.
Essentially, the
mechanism underlying   their success is the  cumulative advantage on
those indexes. However, the effect of cumulative advantage in producing  publications is not
so strong. The tail of the quantitative distribution  of the publications produced by a group of researchers  is much shorter than that of the citation distribution  of those researchers.
Therefore, the prediction methods    of publication productivity would be  different from those   of citation-based indexes and the $h$-index.
\section*{The data}


Due to its nature of regression, the provided  model   needs a training dataset containing  enough productive researchers.  Therefore, the model needs a large training dataset,  spanning a long time interval.
Meanwhile, name ambiguities exist in bibliographic data, which manifest themselves in two ways: one person is identified as two or more entities (splitting error); two or more persons are identified as one entity (merging error)\cite{Milojevic2013,Xie2019merging}. Merging errors would generate    names with a number of publications far more than ground truth, invalidating the prediction results of the model. Therefore, a training dataset with limited errors is required.




The dblp computer science bibliography provides a dataset satisfying above requirements, which consists of the open bibliographic information on major journals and proceedings of computer science (https://dblp.org). The dataset has been  cleaned by several  methods of name
disambiguation and checked manually. For example, the ORCID information has been utilized regularly to correct numerous cases of homonymous and synonymous. We extracted parts of the data at certain time intervals as training   and test datasets  (Table~\ref{tab1}). These parts totally consist of 220,344 publications  in 1,586 journals and proceedings, which are produced by  328,690 researchers at the years from 1951 to 2018.

 Sets 1 and 2 are used to extract the historical publication quantities of test researchers in  Sets 3 and 4 respectively.
Set 5 is  used as a training dataset.
Sets 6 and 7 are used to test the prediction results for the researchers in  Sets 3 and 4 respectively.
 Due to the size and the time span
 of the analyzed dataset, this study would not be treated as a case study.
   The provided model is at least   suitable for the    community of computer science.
Note that the term   ``researcher" in this paper refers to   an author   of the dblp dataset.




\begin{table}[!ht] \centering \caption{{\bf Certain subsets  of the dblp   dataset.} }
\begin{tabular}{l ccccccccccc} \hline
Dataset&  $a$   & $b$ &  $c$  & $d$ & $e$ & $f$    \\ \hline
Set 1 &1951--1995 & 20,781&   20,666& 346& 1.556&1.565\\
 Set  2   &1951--2000 &  38,149& 35,643& 542 &1.571& 1.681\\
   Set   3   &1995 & 3,709  &   2,268& 160  &  1.137 & 1.859 \\
 Set  4  &2000 & 5,741   &   3,600& 257  & 1.184 & 1.888 \\
 Set  5 & 1995--2009 & 87,140&62,636& 931& 1.538&     2.139\\
  Set   6 &  1996--2013 &  116,231& 80,193 & 1,150 &1.557& 2.257 \\
 Set   7 &  2001--2018 &  301,741& 184,701 & 1,495 &1.733& 2.831 \\
\hline
 \end{tabular}
  \begin{flushleft} The index  $a$:    the time interval of data,  $b$: the number  of researchers, $c$: the number of publications,  $d$: the  number of journals,  $e$: the average number of researchers' publications, $f$: the average number of publications' authors. \end{flushleft}
\label{tab1}
\end{table}

 \section*{Motivation} 
\begin{figure*}[h]
\centering
\includegraphics[height=2.3   in,width=4.6   in,angle=0]{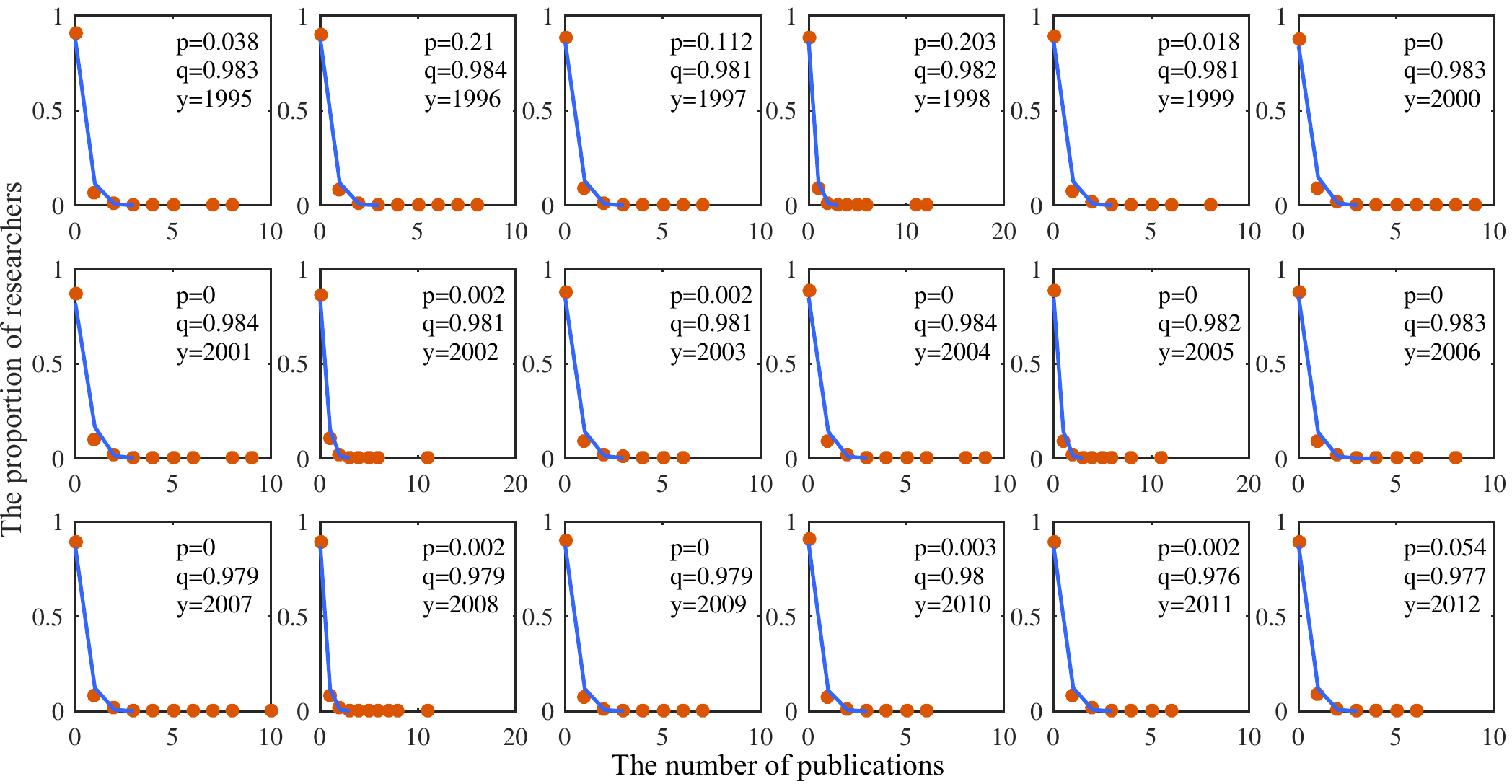}
\caption{    {\bf The quantitative distribution of researchers' publications.  }
Consider the researchers of the training dataset  who have  publications at   $y$
 and    no more than 10 publications at  $[1951,y]$. Index   $q$ is their proportion  to   the total researchers with publications at   $y$.
The KS test   rejects   that
  the quantitative distribution  of  these researchers' publications produced at   $y+1$  (red circles) 
is  a Poisson (blue lines), when $p\leq0.05$. }
 \label{fig1}
\end{figure*}

This study is a data-driven one, inspired by the features of the quantitative distributions of researchers' publications.
Consider the researchers in the training dataset who have 
  publications at the year $y$ and no more than $10$ publications at $[1951, y]$. Consider the quantitative distribution of 
their publications produced at $y+1$, where $y=1995,...,2012$.
 The   Kolmogorov-Smirnov (KS) test rejects to regard them as Poisson distributions because of their tail (Fig.~\ref{fig1}).
 One can also find that  the quantitative distributions of publications produced by   relatively large groups of researchers in the dblp dataset have a fat tail, which  also appears  in other empirical datasets\cite{XieO2016}.


The emergence of their fat tail can be explained  as a result of the cumulative advantage in producing publications or   the diversity of researchers' ability\cite{XieO2018}.
In more detail,
previous studies show that
the   distributions
are featured by a trichotomy, comprising a generalized Poisson head, a power-law middle part, and an exponential cutoff\cite{Xie2019Scientometrics}.
The trichotomy  can be derived from a range of ``coin flipping" behaviors, where the probability of observing ``head" is dependent on observed events~\cite{Consul}.

The event of producing  a publication can be regarded as an analogy of observing ``head", where the probability of publishing is also affected by previous events. Researchers would easily produce  their second publication  compared with their first one. This is a cumulative advantage, research experiences accumulating in the process of producing publications. It displays as the transition from the generated Poisson head to the power-law part. Aging of researchers' creativity is against cumulative advantage, which displays as the transition from the power-law part to the exponential cutoff.

The quantitative distributions of researchers' publications can be fitted by a mixture of Poisson distributions\cite{XieO2016}. Therefore, we could expect to partition researchers into specific  subsets, such that the quantitative distribution of publications produced by the researchers of each subset is a Poisson. When restricting into a short time interval, the effects of cumulative advantage and  aging would be not significant. However, the diversity of researchers in publication history cannot be eliminated only by shrinking the observation window in the time dimension.
Therefore, we provided a split scheme to eliminate the diversity as follows.


Consider  the researchers
 who produced publications  at the two intervals $[T_0,T_1]$ and $[T_1,T_2]$.
Partition the latter one into $J$ intervals with cutpoints $T_1 = t_0 < t_1 <
\cdots < t_J = T_2$.
   The half-closed interval  $(t_{j-1}, t_j]$ is referred to as  the $j$-th time interval, where $j=1,2,...,J$.
  Partition   the researchers with no more than $I$ publications at $[T_0,t_j]$ into  $I$ subsets
  according to  their historical publication quantity at $[T_0,t_{j-1}]$. That is,
  the $i$-th subset consists of the researchers with $i$   publications at $[T_0,t_{j-1}]$.

\begin{figure*}[h]
\centering
\includegraphics[height=2.3   in,width=4.6    in,angle=0]{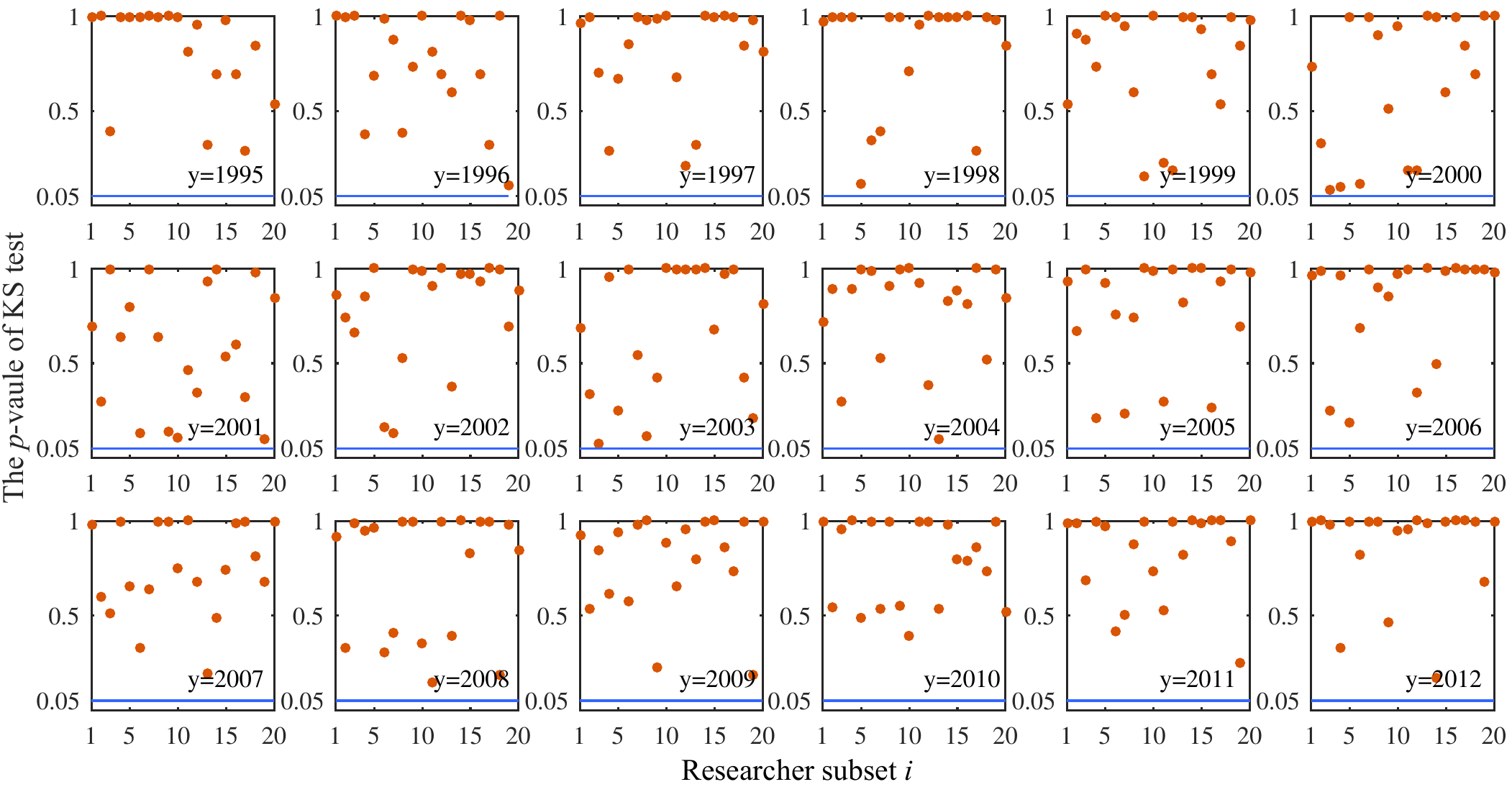}
\caption{    {\bf Eliminating the diversity in   historical   publication quantity induces    Poisson   distributions.  }
 Consider the researchers who produced publications at $y$ and $i$ publications at $[1951,y]$.  The KS test cannot reject
the quantitative distribution of  
the publications produced by these researchers at  $y+1$ is a Poisson, $p$-value$>0.05$. 
 }
 \label{fig2}
\end{figure*}

Let the    $i$-th subset  at the $j$-th time interval  be
the subset of the researchers with $i$   publications at $[T_0,t_{j-1}]$.  
 Fig.~\ref{fig2} shows that  the publication quantity  of a researcher 
    of  the $i$-th subset ($i=1,...,20$) at each    observed time interval ($y+1=1996,...,2013$)  follows a Poisson distribution.  This  inspires us to provide a piecewise Poisson   model.

  

\section*{The piecewise Poisson model}


Our study   considers the creativity  on producing publications,   termed    ``publication creativity".
 The provided model gives a method to  measure it   through researchers'   publication quantity.
 Recall   that  the    $i$-th subset  at the $j$-th time interval  is
the subset of the researchers with $i$   publications at $[T_0,t_{j-1}]$.  
 Denote
  its publication creativity  by $\lambda_{ij}$. Assume  $\lambda_{i1}>0$ and
\begin{equation}\lambda_{ij}=\lambda_{i1}   \mathrm{e}^{{ \beta_i} ({t}_j-t_1)  }, \label{eq1}
\end{equation}
   where  $\beta_i $ tunes the  effect of  time  ${t}_j$.  Given   $i$,
the formula in Eq.~(\ref{eq1}) is exactly  the Poisson   model   (see its definition in Appendix A),
because the quantity of the  publications produced by  a researcher in the $i$-th subset   at $(t_{j-1},t_j]$
  follows a Poisson distribution (Fig.~\ref{fig2}).
Therefore,  we named the provided model piecewise Poisson model.  
 





Now let us show 
the calculation of  publication creativity.
Consider a   dataset
consisting of the researchers having publications at the  time interval $[T_0, t_{L-1}]$  and their publications at the time interval $[T_0, t_{L}]$, where $1<L\leq J$.
The  publication quantity of a researcher at $(t_{j-1},t_{j}]  $ is his or her publication quantity at that time interval. Then, we
defined the   publication productivity of the $i$-th subset 
 in the   dataset
 at  $(t_{j-1},t_{j}]  $ to be its average publication quantity at that time interval, and denote it by $\eta_{ij}$.
   It 
can be calculated  as \begin{equation}\eta_{ij}=\frac{m_{ij}}{ n_{ij}}, \label{eq2}
\end{equation}
  where  
   $n_{ij}$ is the number of the researchers with $i$ publications at $[T_0,t_{j-1}]$, and
   $m_{ij}$ is the
   number of publications produced by those  researchers  at  $(t_{j-1},t_j]$.

\begin{figure*}[h]
\centering
\includegraphics[height=2.2 in,width=4.5      in,angle=0]{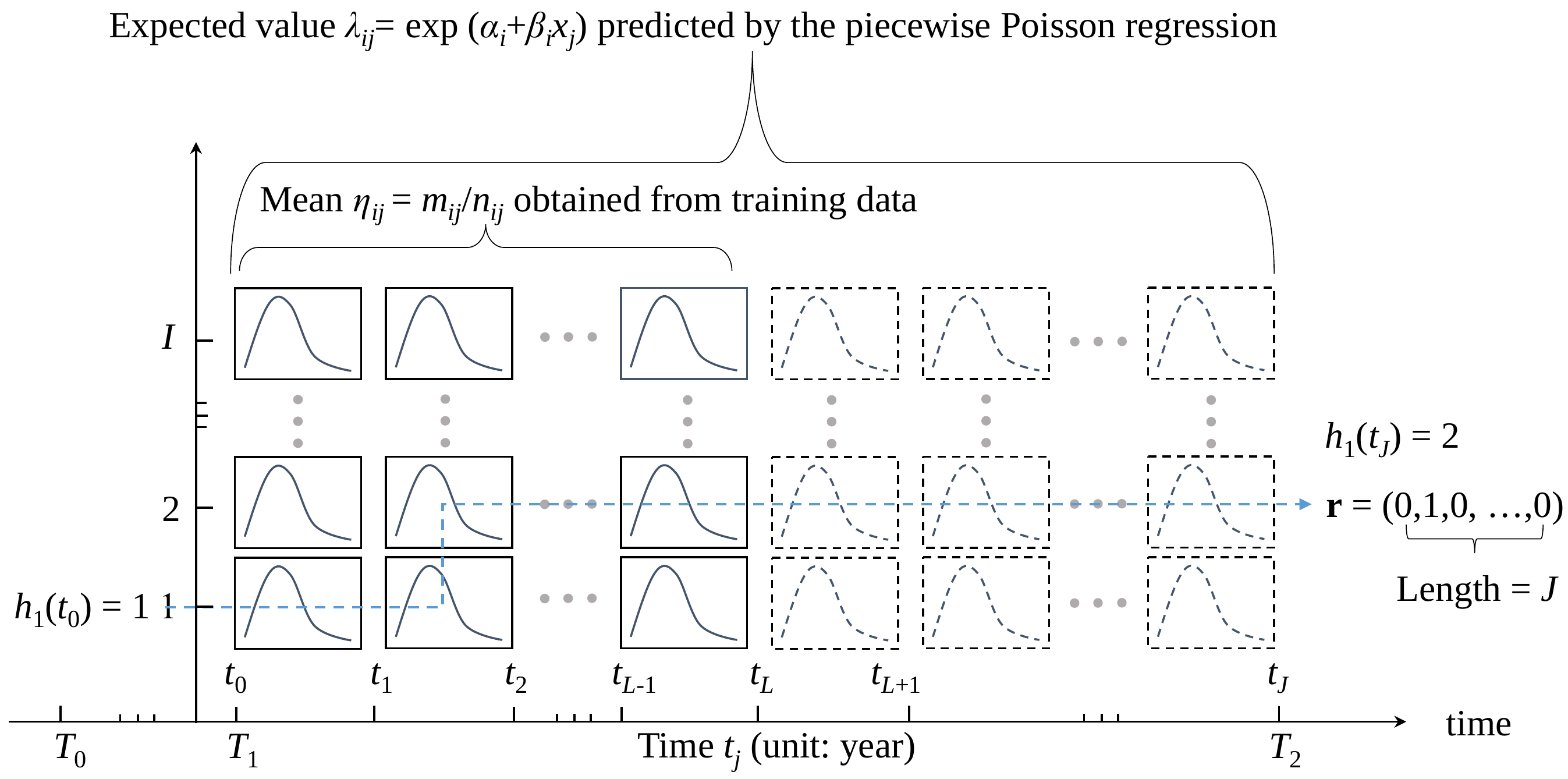}
\caption{   {\bf   An illustration of the  piecewise Poisson model.} A training dataset is used to calculate the publication productivity $\eta_{ij}$.
The publication  creativity  $\lambda_{ij}$  is calculated  by our model. Vector ${\bf r}$ records the predicted   publication quantity   of a researcher   at  each time interval  $(t_{j-1},t_j]$, where $j=1,...,J$. }
 \label{fig3}
\end{figure*}

 Define
the publication creativity $\lambda_{ij}$ to be the expected value of $\eta_{ij}$. Therefore, we need a training dataset to  calculate  $\lambda_{ij}$ by regression.
  Taking logs in Eq.~(\ref{eq1}) and
substituting $\eta_{ij}$ into it,
we obtained
\begin{equation}\log \eta_{ij}= \alpha_i+\beta_i (t_j-t_1),\label{eq3}
\end{equation} where $\alpha_i=\log \lambda_{i1}$, and   $j=1,...,L$.
  The linear regression is utilized   to calculate $ \alpha_i$ and $\beta_i$. Eq.~(\ref{eq3}) describes  the relationship between   $\eta_{ij}$ and   $t_j$ given  $i$.
The relationship is significant for the majority researchers in the training  dataset used here; thus we can let 
 $\lambda_{ij}=\mathrm{e}^{\alpha_i+\beta_i (t_j-t_1)}$.  Fig.~\ref{fig3} shows an illustration  of   the provided model.

 Algorithm~\ref{tab2} is provided to  predict researchers'  publication quantity    at the time interval  $[t_X,t_Y]$, where $t_0\leq t_X<t_Y\leq t_J$.
  Denote  the publication quantity   of researcher   $s$   at  $[T_{0},t_l]$ by $h_s(t_l)$.
The algorithm gives  $h_s (t_Y)$  the predicted  publication quantity of researcher $s$ at    $[T_0, t_Y]$.
  Due to its regression nature,  the algorithm cannot exactly predict the publication quantity  for  an individual, but it can be suitable  for a group of researchers.

 Note that the training dataset would be not enough for using linear regression given a large    $i$. Therefore,  the model cannot be applied to productive
 researchers with a publication quantity   more than $I$.
Our model can only be used to predict     publication quantity  for
the researchers with     a  publication quantity at $[T_0,t_X]$   no more than  a given integer $I_1<I$.



\begin{algorithm}
\caption{Predicting researchers' publication  quantity.}
\label{tab2}
\begin{algorithmic}
\REQUIRE ~~\\ 
the $h_s(t_X)$ of any  test researcher $s$;\\
the parameter $J$; \\
the  matrix  $(m_{ij}/n_{ij})_{I\times L}$.
\ENSURE ~~\\ 
  the predicted productivity  $h_s (t_J)$ of   researcher $s$.

\FOR{$i$ from $1$ to $I$}
\STATE{calculate $ \alpha_i$ and $\beta_i$ in  Eq.~(\ref{eq3}) by the  linear regression; }
\STATE{let  $\lambda_{ij}=  \mathrm{e}^{\alpha_i+{ \beta_i} ({t}_j-t_1)}  $ for $j=1,...,J$;   }
\ENDFOR

\FOR{each    researcher $s$}
\STATE{initialize $h=h_s(t_X)$;}
\FOR{$l$ from $X+1$ to $Y$}
\STATE{sample an integer  $r$ from   Pois$(\lambda_{hl})$; }
\STATE{let  $h= h    + r$;}
\ENDFOR
\STATE{let $h_s (t_Y)=h$.}
\ENDFOR
\end{algorithmic}
 \end{algorithm}

\section*{Experiments}
Now the model is applied to the dblp dataset. The training dataset    consists of the researchers  in Set 5 and their historical publication quantity  from Set 1,
  Its parameters  are   $I=40$,   $J=23$,   $L=14$,  $T_0=1951$,  $T_1=t_0=1995$,  $t_L=2009$,  and  $T_2=t_J=2018$. 
 The test dataset    consists of the researchers  in Set 4, their historical publication quantity  from Set 2,
 and their annual publication quantity  from Set 7.       Its parameters  are 
  $t_X=2000$,  and $t_{Y}=2018$.

The matrix of  publication productivity   $(m_{ij}/n_{ij})_{I\times L}$   is   calculated  based on   the training dataset.
For example,
 $n_{11}$  is the number of researchers with one publication at  $[1951,1995]$, and
 $m_{11}$ is the publication quantity  of those researchers at the year $1996$. Then,   
  $\alpha_i$ and $\beta_i$ are calculated   by applying the linear  regression to Eq.~(\ref{eq3}).

\begin{figure*}[h]
\centering
\includegraphics[height=2.3   in,width=4.6     in,angle=0]{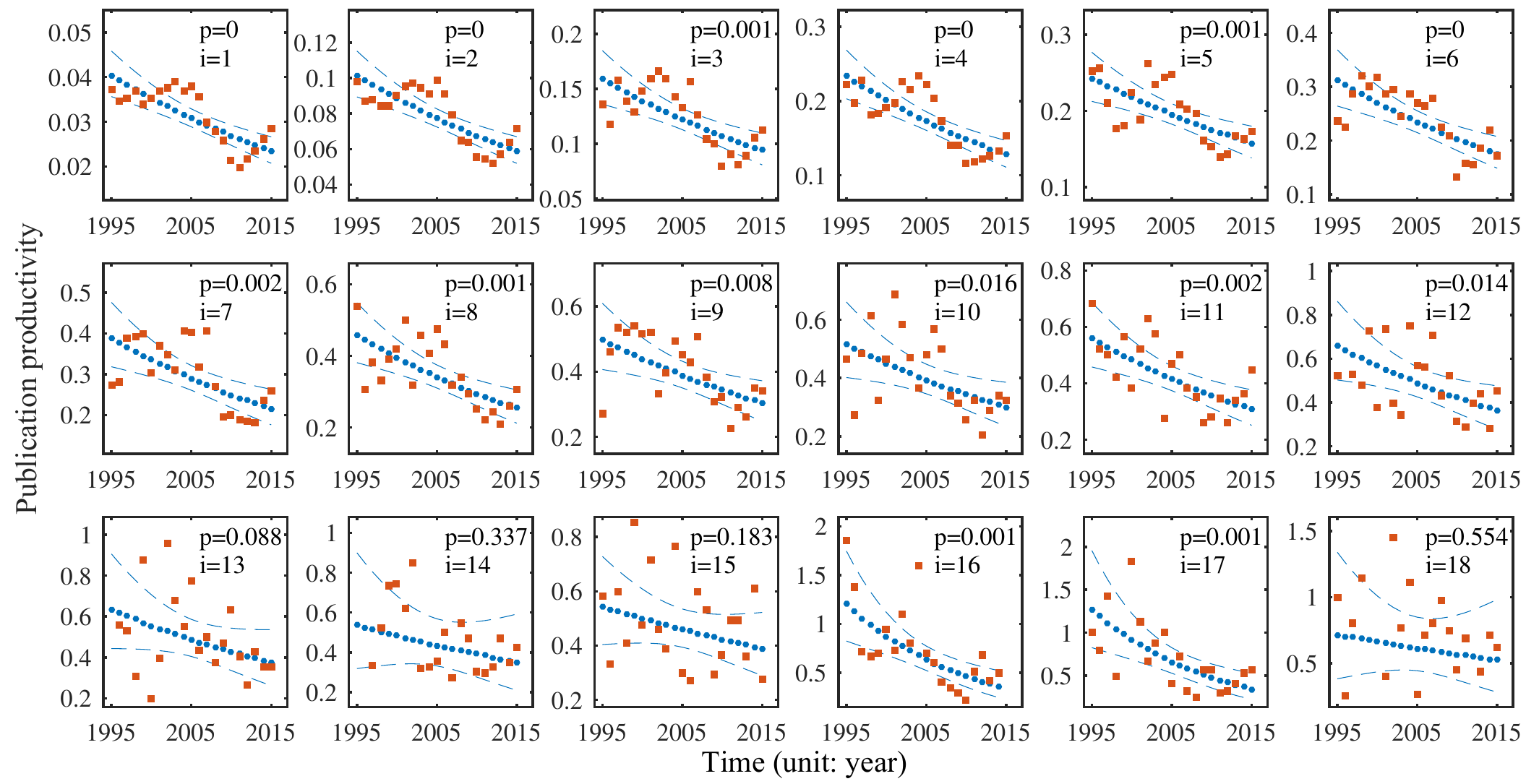}
\caption{   {\bf  The relationship between    publication productivity   and    time.}
 Consider the researchers    who have $i$  publications at $[1951,y]$. Calculate   their  publication productivity  at  $y+1$ (red squares).
 The solid dot lines show the predicted results by the  Poisson regression, and the dashed lines show confidence intervals. 
 The relationship   is significant when $p<0.05$.  }
 \label{fig4}
\end{figure*}

 The $\chi^2$ test  indicates that  $ \eta_{ij}$ significantly correlates to $t_j$ given   $i\leq12$ (see the $p$-values in  Fig.~\ref{fig4}). 
 That is, the significance holds  for   $99.5\%$ researchers in the training dataset.
 Thus, we can let $\lambda_{ij}=\mathrm{e}^{\alpha_i+\beta_i (t_j-t_1)}$. 
 In the experiment here,
  we can only predict   publication quantity  for  98.76\%  researchers of the test  dataset (who have  no more than $I_1=13$ publications at $[T_0,t_X]$) due to    the  maximum  publication quantity of our model. Two methods are provided to testify  the  effectiveness of the model as follows.

Firstly,    we tested the model   by its prediction on  the
   evolutionary  trend  of researchers'  publication quantity.
     Consider the test researchers   who produced $i$ publications at  $[T_0,t_X]$.
 Let $n(i,y)$ be the average   publication quantity of these researchers  at   $[1951,y]$, and $m(i,y)$ be the predicted one.
  Fig.~\ref{fig6} shows their   trend   about $i$  given
   $y$.
  The correlation between them is measured by the  Pearson  correlation coefficient\cite{Hollander} on individual level ($s_1$) and that on
 group level ($s_2$). Index $s_1$ decreases over time, whereas $s_2$ keeps high.
  It indicates that  the model is unapplicable to    the long-time prediction for an individual, but can be applicable for a group of researchers.



  \begin{figure*}[h]
\centering
\includegraphics[height=2.3   in,width=4.6    in,angle=0]{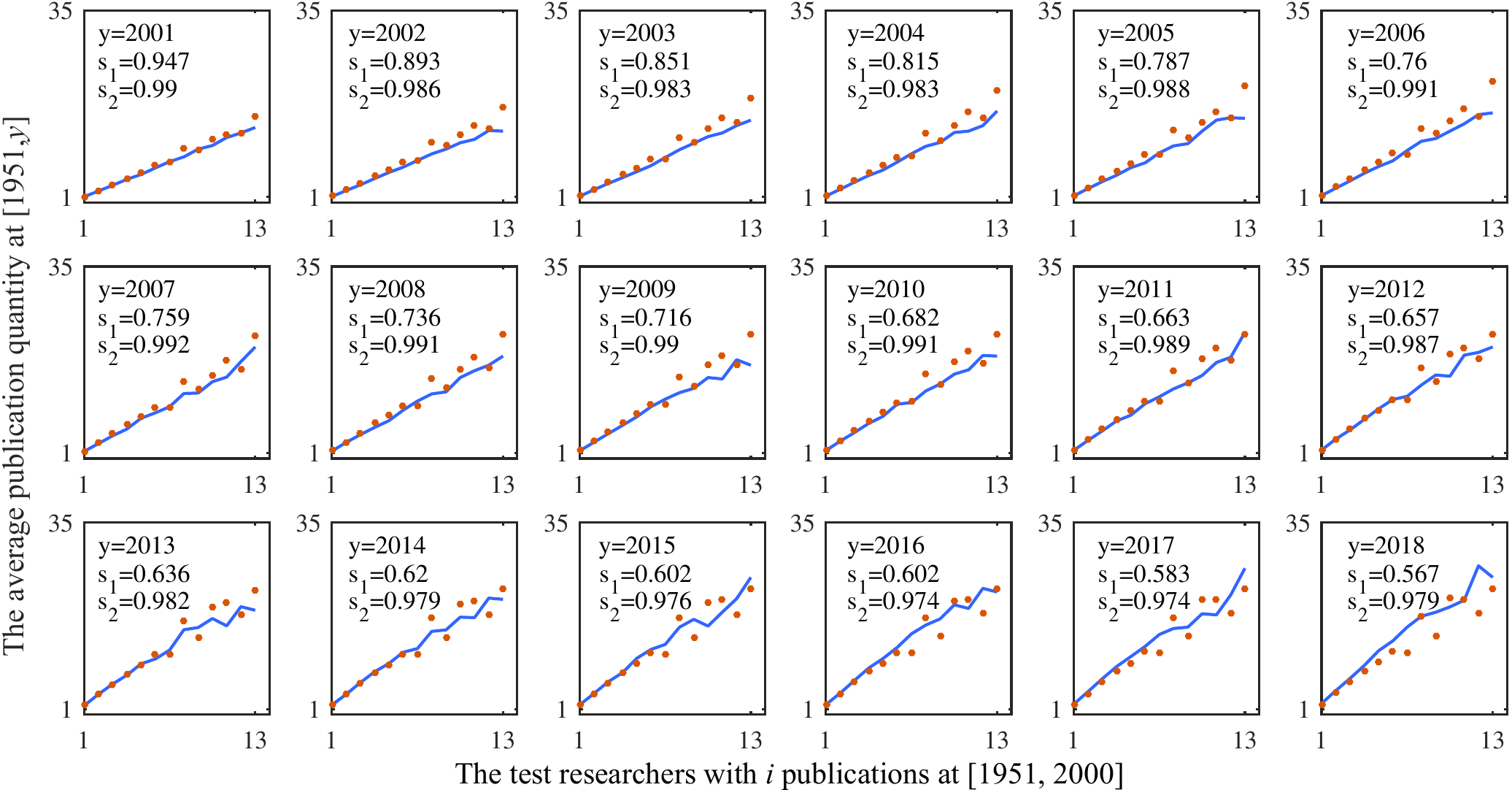}
\caption{   {\bf  Fittings on the evolutionary trend of   researchers' publication quantity.} Consider  the test researchers who have $i$ publications at $[1951,2000]$.
 Panels show 
  the average number of publications produced  by  these researchers at    $[1951,y]$ ($n(i,y)$,  red dots) and the predicted one    ($m(i,y)$, blue lines).  Index $s_1$ is the  Pearson  correlation coefficient
     calculated
 based on  the  list of researchers'   publication quantity and that of their predicted one. 
   Index $s_2$ is that based on the sorted 
    lists.}
 \label{fig5}
\end{figure*}

Secondly,   we tested  the model   by its prediction on  the quantitative distribution of researchers' publications.
 We  compared the   distribution for the publications produced by   the test researchers at $[T_0,y]$   with the    predicted  one.
Fig.~\ref{fig6} shows that
a fat tail emerges in the evolution of
the  ground-truth distributions, because a small fraction  of researchers produced  many  publications. However, our model cannot predict over-exaggerated productivity  due to the  maximum  publication quantity that can predicted by our model.   Therefore, the KS test rejects that the compared  distributions are the same with the growth of time (see the $p$-values in Fig.~\ref{fig6}), although there is a coincidence in their forepart. It indicates that the prediction precision   for productive researchers  needs to be improved.







  \begin{figure*}[h]
\centering
\includegraphics[height=2.3   in,width=4.6    in,angle=0]{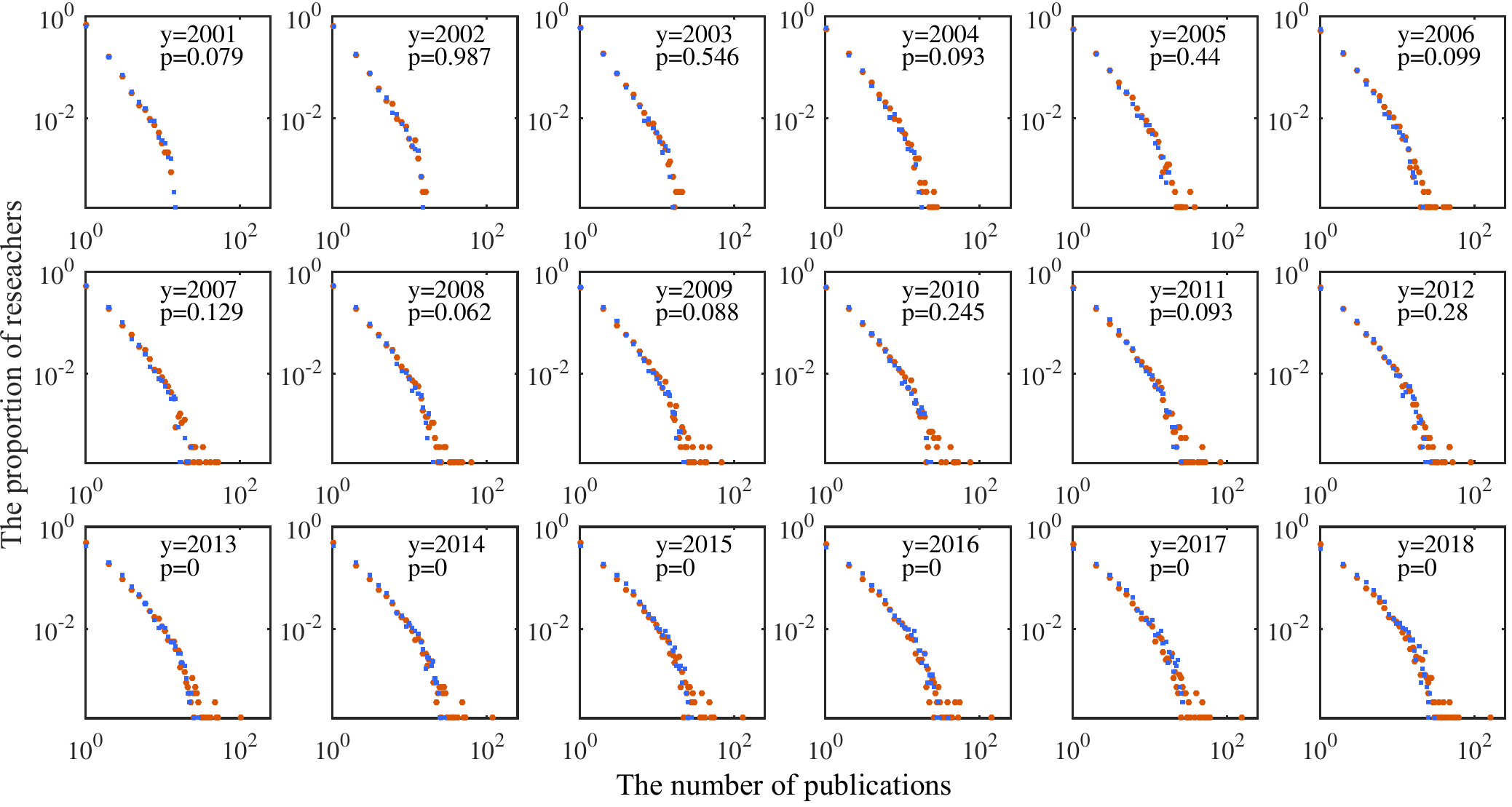}
\caption{   {\bf  Fittings  on   the quantitative distribution of researchers' publications.}
Panels  show this distribution for the publications
produced by the test  researchers   at  the time interval $[1951 ,y]$ (red circles) and
     the predicted one (blue squares). When  $p>0.05$, the KS test cannot reject the null hypothesis that the compared distributions are the same.  }
 \label{fig6}
\end{figure*}

\section*{Comparisons with previous results}

\begin{figure*}[h]
\centering
\includegraphics[height=1.8    in,width=2.5    in,angle=0]{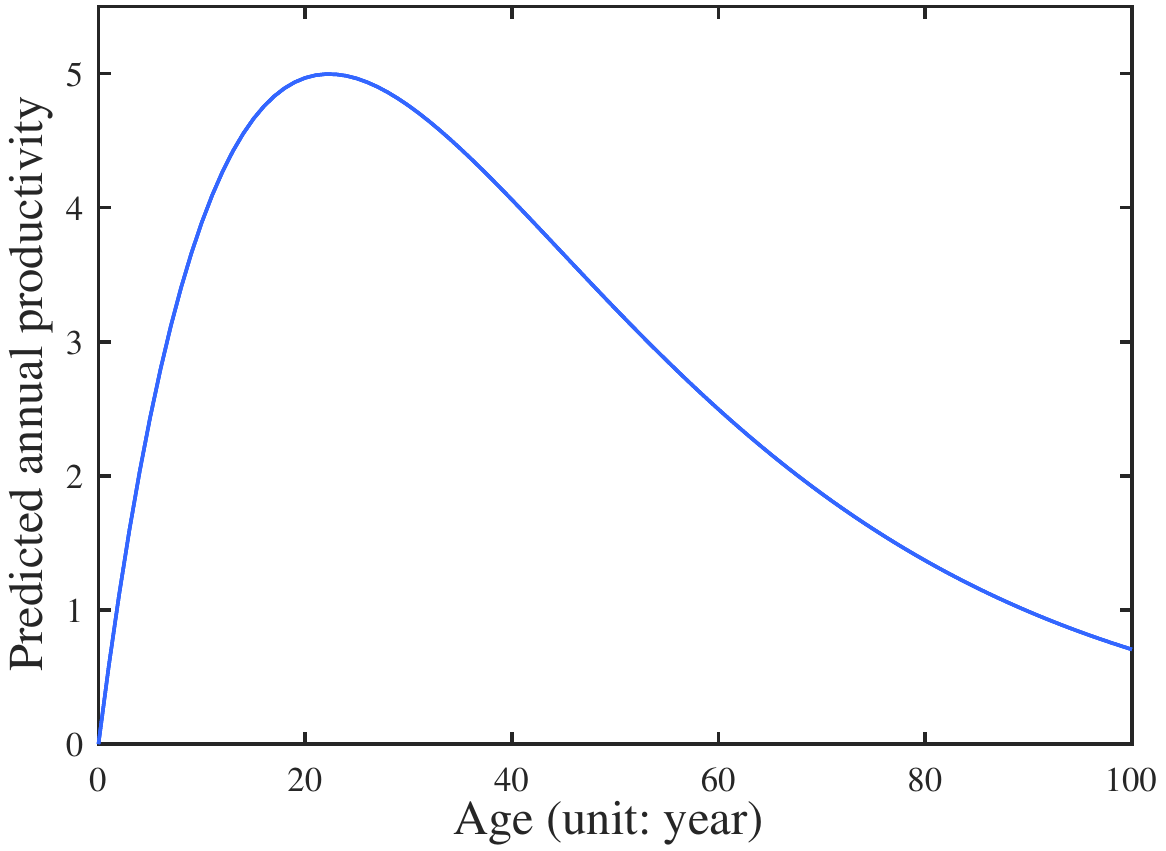}
\caption{   {\bf The publication productivity  predicted by the formula in Eq.~(\ref{eq7}).}  Panel show the curve of this formula with
 the parameters provided by Simonton: $a = 0.04$, $b = 0.05$, and $c = 61$.}
 \label{fig7}
 \end{figure*}

Firstly, we discussed the possibility  of utilizing the prediction  formula  provided by Simonton\cite{Simonton1984}:
\begin{equation}
p(t)=c(\mathrm{e}^{-at}-\mathrm{e}^{-bt}),\label{eq7}
\end{equation}
where  $a,b,c\in \mathbb{R}^+$. Parameter  $a$ is termed
the   ``ideation rate",   $b$ is termed
the  ``elaboration rate", $c=abm/(b-a)$, and
$m$   represents the maximum number of publications  a researcher can produce in his  lifespan.
This formula   theoretically    expresses a researcher's publication  productivity     by a function of
time $t$.
 With the parameters  in Reference\cite{Simonton1984},  the shape of this   curvilinear function  is shown in Fig.~\ref{fig7}.    

As aforementioned, the cumulative advantage and the aging of creativity have impacts on researchers' publication productivity. One can think that a researcher's publication productivity is proportional to   his publication quantity  in his early age of research. The more publications he has, the higher his publication productivity. As his age increases, his creativity decreases and will dry up in his later period of research.   The formula in
Eq.~(\ref{eq7})    expresses  this    evolution of publication productivity.

\begin{figure*}[h]
\centering
\includegraphics[height=2.3   in,width=4.6    in,angle=0]{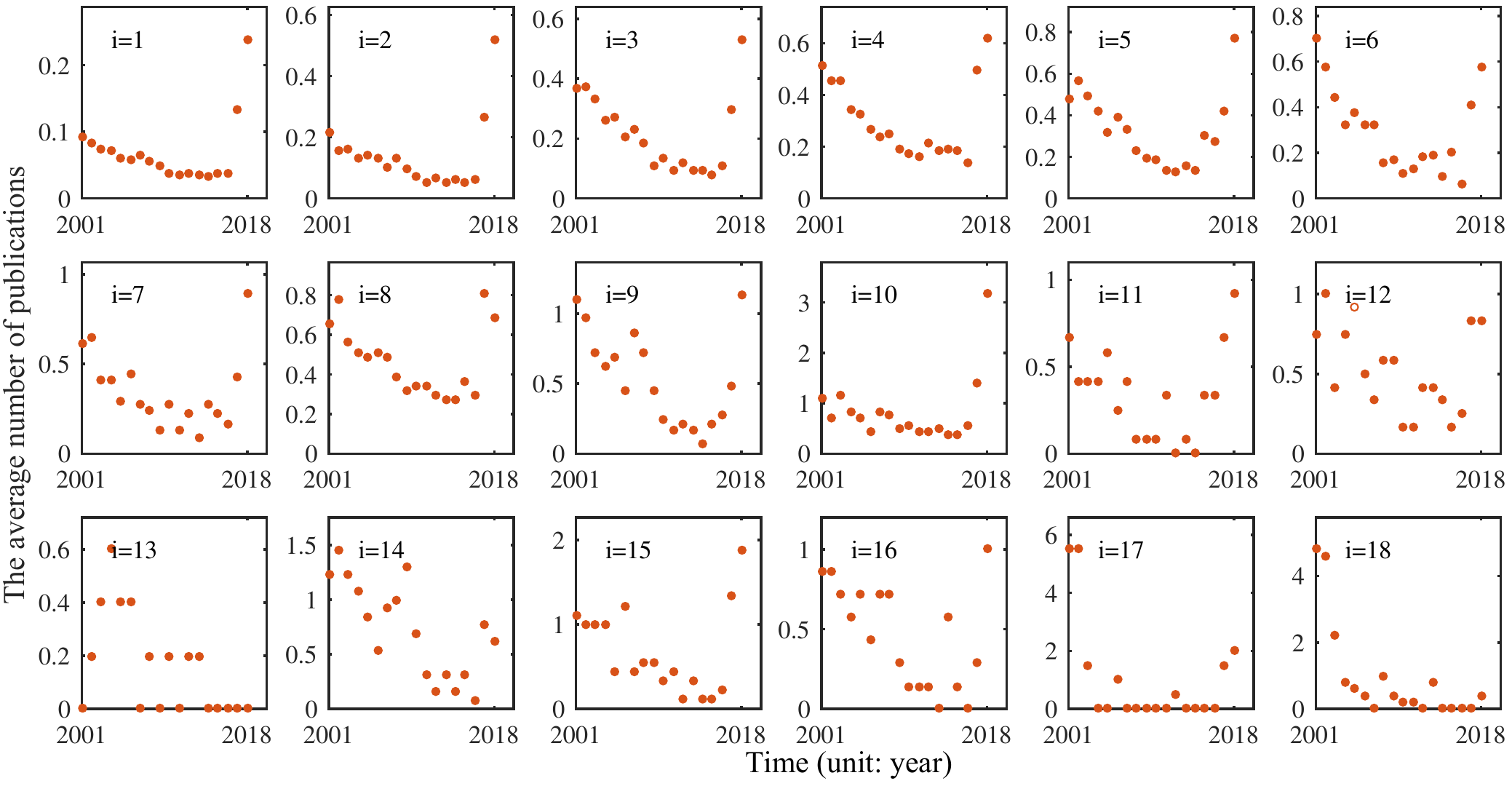}
\caption{   {\bf  The average annual publication quantity  of   test researchers.} Consider
 the  test researchers with  $i$ publications at  $[1951,2000]$.
The red circles show the average   publication quantity of these researchers  at each  year from $2001$ to $2018$. 
 }
 \label{fig8}
\end{figure*}

Consider  the test researchers with  $i$ publications at $[T_0,t_X]$, where $i=1,..,18$.  
Fig.~\ref{fig8} shows the average  publication quantity  of these researchers  at each year from $2001$ to $2018$, which cannot be fitted by the formula in Eq.~(\ref{eq7}). One possible explanation of this inconformity is the variation of the personnel structure on the researchers who produce publications. Note that the formula is provided at the year 1984. In recent thirty years, the number of academic masters and doctors dramatically increases. They contribute a large fraction of publications during their study periods. Many of them will not do research after graduation, and thus will not continue to produce publications. Therefore, the formula in Eq.~(\ref{eq7}) is unsuitable for describing the evolutionary process of their productivity. Meanwhile, it can be suitable for the  fruitful scientists who have a long research career. However, in the dblp dataset, the number of these researchers is quite small, because more than $99.5\%$  researchers produced  no more than $40$ publications.





Secondly, we discussed the practicability of utilizing
a researcher's historical
publication productivity to predict  his or her future productivity.
Therefore, we should know that is there any
predictable components of   the productivity that can be extracted via
autoregression. Previous studies found that
those components are  significantly positive predictors to citation-based indexes and $h$-index,    constituting  the
principle terms of
the prediction  formulae  of those indexes.


The  mechanism
of generating these
autoregressive
components
  is the  cumulative advantage in receiving  citations and in the evolution of $h$-index. Previous  empirical studies show that the number of citations received in the future depends on the number of citations already received\cite{Price1}. However, the effect of  cumulative advantage in producing publications   is much weaker than that in receiving citations. It is reflected by the short tail of the quantitative distribution  of a group of researchers' publications, compared with that of the citation distribution of    the same researchers\cite{XieX2017}.


 In statistics,   autoregressive models specify that the response variable depends linearly on its previous values with a stochastic term. The  advantage of those model is that not much information is   required,   only the self-variable series.
 If the autocorrelation coefficients
 of the response variable series
 are   smaller than 0.5,   autoregressive models are not suitable for prediction task. That is,  the coefficients of
autoregressive components are not large enough   to be significant  predictors.

 The autocorrelation coefficient  of ${\bf y}=(y_1,...,y_T)$ with lag $l$ (where $l<T$) is defined  as
\begin{equation}
 r_l=    \frac{ \sum^{T-l}_{t=1}(y_t-\bar{y})(y_{t+l}-\bar{y})} { \sum^{T-l}_{t=1}(y_t-\bar{y})(y_{t+l}-\bar{y})},\label{eq7}
\end{equation}
where $\bar{y} $ is the mean of ${\bf y}$'s elements\cite{Box1994}.
  We constructed a  time series   ${\bf y}_s=(y_s(t_0),...,y_s({t_J}))$ to record the quantitative information of  publications for a researcher $s$, where
 $y_s(t_j)$ is the number of his   publications produced at $[T_0,t_j]$  for $j=0, ..., J$.

 We calculated the autocorrelation coefficients of ${\bf y}_s$ for any  researcher $s$
 in the  test dataset.  We found that   these coefficients are almost smaller than 0.5 given a lag$>1$ . Therefore,  the  historical   publication productivity  of an individual is    not sufficient to predict  his or her future productivity. It indicates that the autoregressive models  may not be suitable  for the prediction of publication productivity; thus the schemes of those successful
prediction methods about citation-based indexes and the $h$-index may also be unfeasible.

  \begin{figure*}[h]
\centering
\includegraphics[height=2.3   in,width=4.6    in,angle=0]{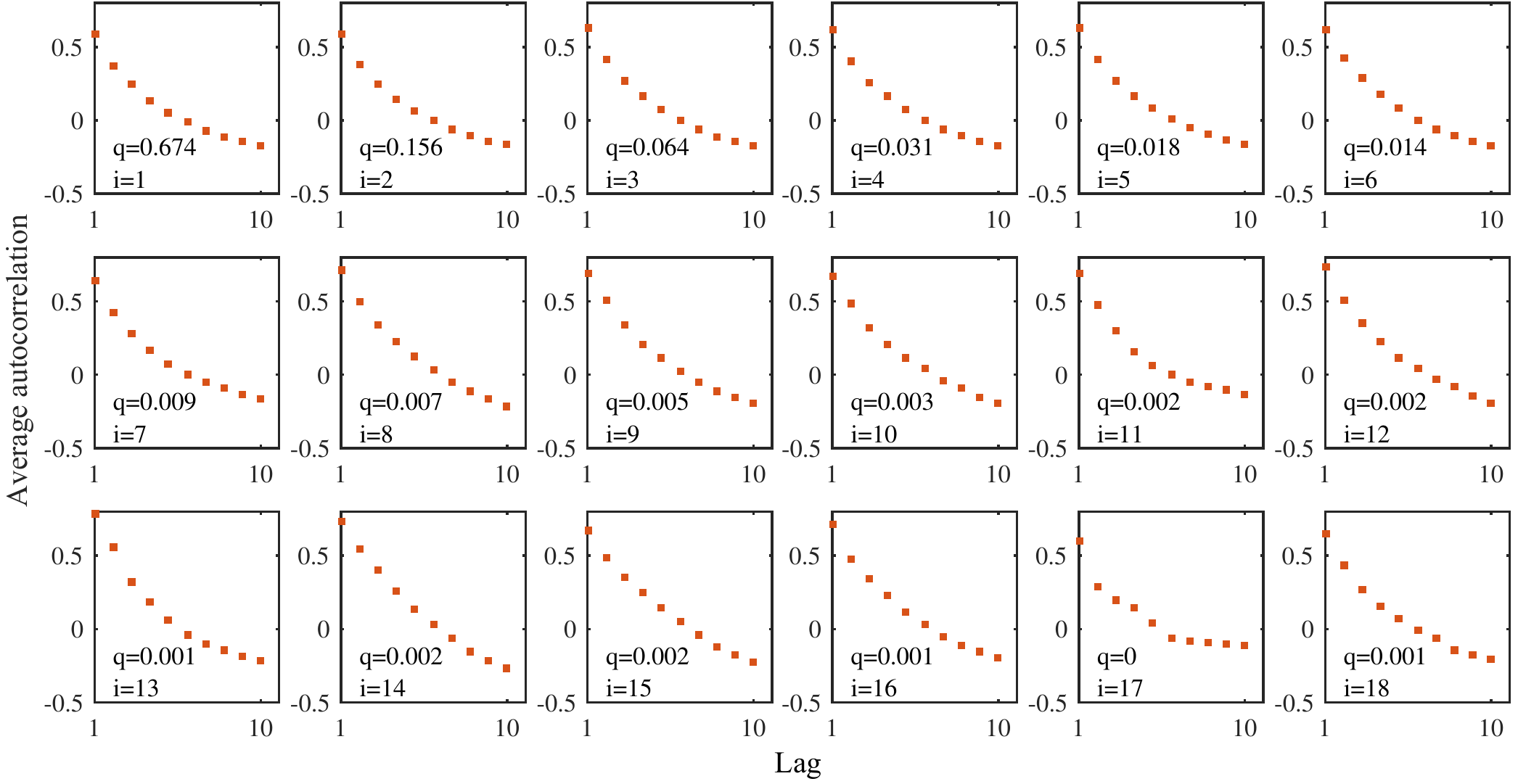}
\caption{   {\bf     Autocorrelation coefficients
 of  the   series on  cumulative  publication quantity.}   Consider this    series of   test researchers from $2000$ to $2018$.
 Panels show the average  autocorrelation coefficients    of the  test  researchers with $i$ publications at $[1951,  {2000}]$. Index $q$ is their proportion to all of the test researchers.}
 \label{fig9}
\end{figure*}




\section*{Discussion and conclusions}
We provided a model to predict the publication  productivity of researchers. The model needs a large training dataset, but there is not much information about researchers required,   only their publications' production time. The model's practicability is testified by the dblp dataset, which exhibits its ability in the prediction of  publication  productivity by the  fine   fittings on the
evolutionary trend of researchers' productivity and the quantitative distribution of their publications.
Due to its nature of regression,  our model has the potential to be extended to assess  the confidence level of prediction results, and thus has clear applicability to empirical research.
  With its prediction results unbiased,  the model may be useful for funding agencies to evaluate the possibility of applicants to complete the quantitative index of publications in their applications.


Our model offers convincing evidence that the publication patterns of many researchers are characterized by a piecewise Poisson process. Even where our model does not provide an exact productivity prediction for  an  individual, it may still be of use in its ability to provide a satisfactory prediction for a group of researchers on average. The prediction results
of our model
offer some comfort by showing that the future of a   group  of researchers  is not so random. The occasional rejection of a paper may feel unjust and indiscriminate, but for a group, such factors seem to average out, rendering the trajectories of researchers' publication productivity  relatively predictable.

Analyzing massive data  to track scientific careers would   help to advance our understanding of how researchers' productivity evolves. 
 The prediction precision   of the model would be improved by utilizing more features of researchers, such as the network features of their coauthorship (degree, betweenness, centrality, etc.), because previous  studies showed that research  collaboration  contributes to scientific productivity\cite{ Lee2005,DuctorL2015,QiMZengA2017}.
However, little is known about the mechanisms governing the evolution   of  researchers' publication productivity.
Predicting the   productivity of an individual  would not be done only by regression as this study did for a group of researchers, due to the randomness of an individual's research. The randomness is displayed in
this study    by the relatively small   autocorrelation coefficients of the time series on a researcher's cumulative publication productivity.
 Therefore, advanced algorithms are needed to synthetically analyze   more features of researchers, such as     journals' annual issue volume, impact factors, and language.

 \section*{Acknowledgments} The author thanks   Professor Jinying Su in the
National University of Defense Technology for her helpful comments
and feedback. This work is supported by the   National Natural   Science Foundation of China (Grant No. 61773020) and  National Education Science Foundation of China (Grant No. DIA180383).

\section*{Appendix A: The Poisson model}
The Poisson model is   a   generalized linear model  of regression analysis\cite{Nelder-Wedderburn1972}. It is used to model count data and contingency tables, thus has potential  to  predict publication productivity.
 The
Poisson model assumes the response variable $y$ follows a Poisson distribution, and assumes the logarithm of its expected value can be expressed by a linear combination of covariates. Let
   $\mathbf {x} \in \mathbb {R}^{n}$ be a vector of   covariates, and $\Phi \in \mathbb {R} ^{n} $ be a vector of
 the covariates' effect. The Poisson model takes the form
\begin{equation}{ \log( {\mathbb{E}} (y|\mathbf {x} ))=\alpha +\Phi \cdot \mathbf {x}},\label{eq4}
\end{equation} where $\alpha\in \mathbb {R}$, and ${\mathbb{E}} (y|\mathbf {x} )$ is the conditional expected value of $y$ given $\mathbf {x}$.
 Note that  Eq.~(\ref{eq1}) can be generalized to deal several  characteristics varying with $i$, namely a vector  ${\bf x}_{ij}$. This study only considered the simplest case: one
characteristic    $t_j$.
\section*{Appendix B: The piecewise exponential model}

%
%

The formula  of the provided model is similar to  that of the piecewise exponential model in survival analysis, which is defined as follows\cite{Cox1972}.
 Assume  that the duration  $t$ of an event is a continuous random variable with probability density function $f(t)$. Let
 $F(t) = \int_{\tau < t} f(\tau)d\tau $, which is the cumulative distribution function. It is the probability that the event has occurred at duration $t$.
 The
survival function  is defined as $S(t)=1-F(t)$, and   the hazard
function   $\lambda(t)=f(t)/S(t)$.


Let $\mathbf{x}_i $ be a vector of   covariates for individual $i$, and
$ \Phi  $ be the vector of covariates'  effect.
   The hazard function at   $t$ for   individual $i$
is assumed to be
\begin{equation}\lambda_i(t, {\bf x}_i  )=\lambda_0   (t) \mathrm{e}^{{\bf x}_i \cdot{\Phi } },\label{eq5}
\end{equation}
where $t\in [0,T]$, and  $\lambda_0(t)$ is a baseline hazard function that describes the risk for individual $i$ with $\mathbf{x}_i  = 0$, and $\mathrm{e}^{{\bf x}_i  \cdot  \Phi  }$ is the relative risk.

Subdivide time into reasonably small intervals and assume that the baseline hazard is constant at each interval,
leading to a piecewise exponential model
\begin{equation}\lambda_{ij}=\lambda_j  \mathrm{e}^{{\bf x}_i\cdot  \Phi  },\label{eq6}
\end{equation}
where $\lambda_{ij}$ is the hazard corresponding to individual $i$ at time $j$, $\lambda_j$ is
the baseline hazard at  $j$.
     Write  $ \Phi$ as  $\Phi_j$ and ${\bf x}_{i}$ as ${\bf x}_{ij}$ to allow for a time-dependent effect of the predictor vector.
Then, we would write
\begin{equation}\lambda_{ij}=\lambda_j  \mathrm{e}^{{\bf x}_{ij}\cdot  \Phi_j  },\label{eq6}
\end{equation}
where is the formula of  the piecewise exponential model.

Although Eq.~(\ref{eq1}) and Eq.~(\ref{eq6}) are similar in   formula,   they  are essentially different.
  The index $j$ in  Eq.~(\ref{eq1}) is   the time, and index $i$ is about researcher subset. The regression is used to calculate
 $\lambda_{i1}$ and  $\beta_i$, which vary with researcher subset $i$ and are free of the index of time $j$. But in Eq.~(\ref{eq6}), the baseline $\lambda_j$ and  the effect $\Phi_j$ are free of $i$ but depend  on   $j$.

\section*{Appendix C: An other example}

\begin{figure*}[h]
\centering
\includegraphics[height=2.3   in,width=4.6     in,angle=0]{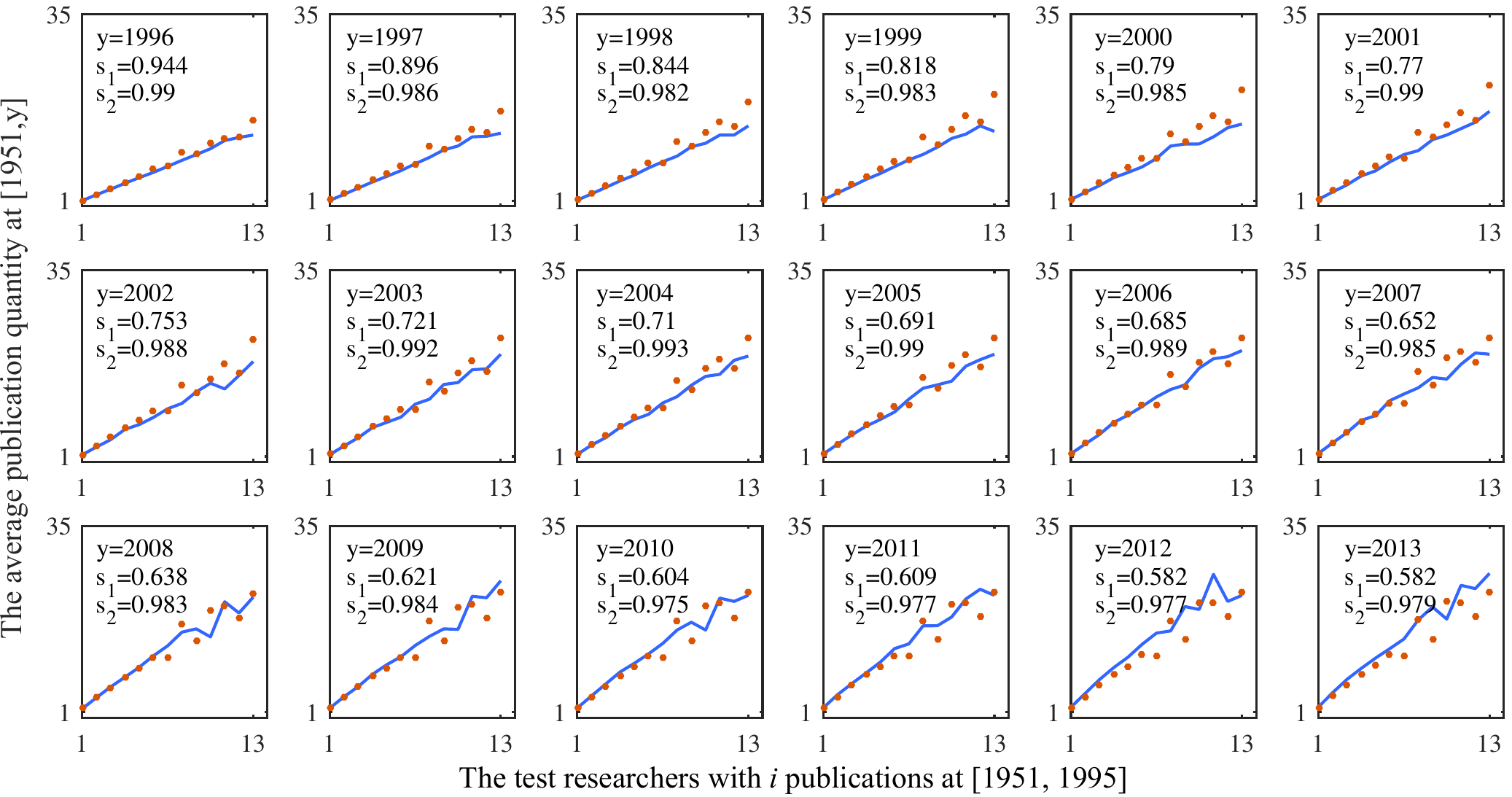}
\caption{   {\bf  Fittings on the evolutionary trend of  researchers'  publication quantity.} Consider  the test researchers who have $i$ publications at $[1951,1995]$.
  Panels show
  the average number of publications produced  by  these researchers at    $[1951,y]$ ($n(i,y)$,  red dots) and the predicted one    ($m(i,y)$, blue lines).  Index $s_1$ is the  Pearson  correlation coefficient
     calculated
 based on  the  list of researchers'   publication quantity and that of their predicted one.
   Index $s_2$ is that based on the sorted
    lists.}  
    \label{fig10}
\end{figure*}

\begin{figure*}[h]
\centering
\includegraphics[height=2.3   in,width=4.6     in,angle=0]{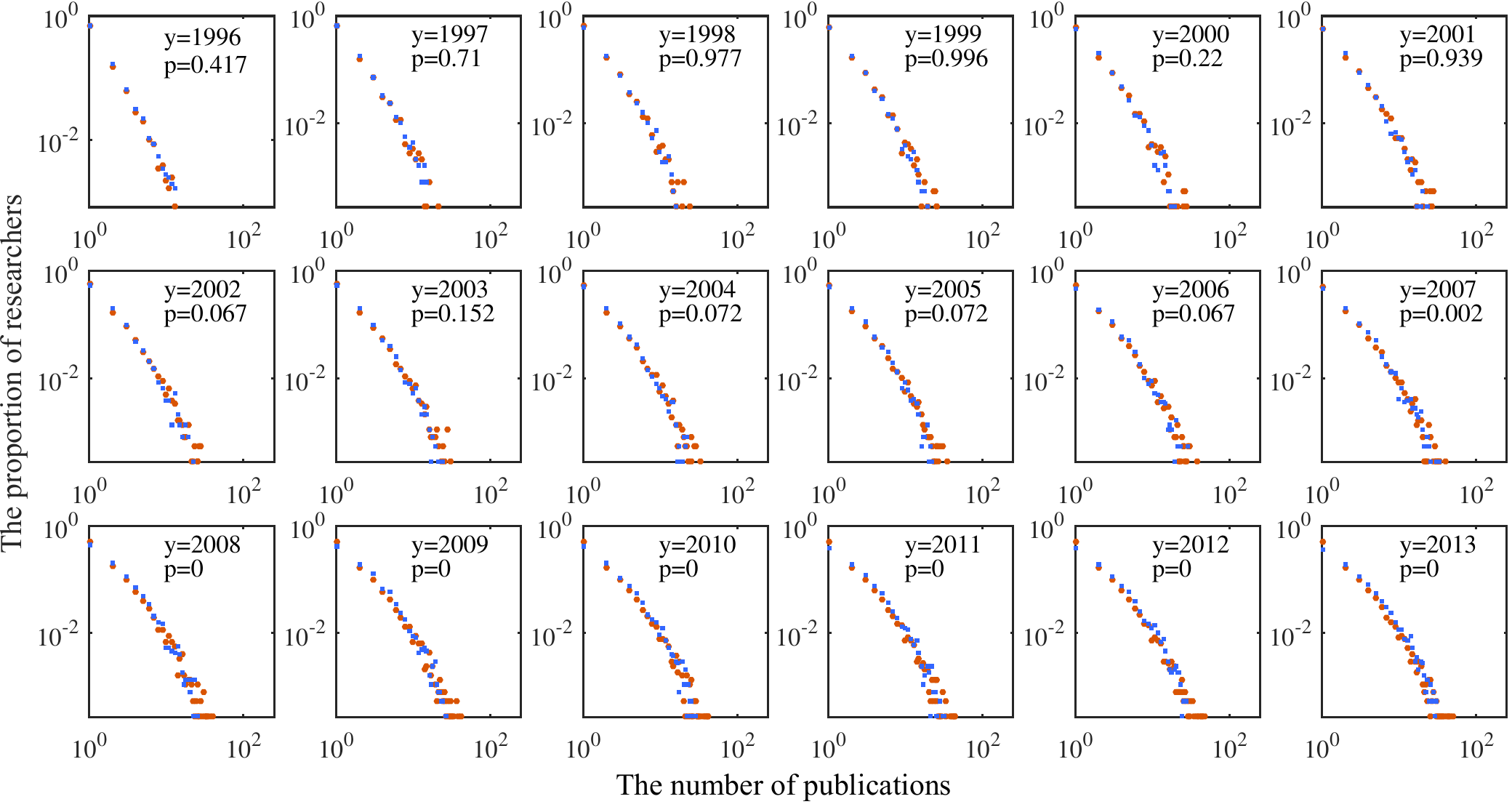}
\caption{   {\bf  Fittings  on   the quantitative distribution of researchers' publications.}
 Panels  show this distribution for the publications
produced by the test  researchers   at   $[1951 ,y]$ (red circles) and
     the predicted one (blue squares). When  $p>0.05$, the KS test cannot reject the null hypothesis that the compared distributions are the same.  }
 \label{fig11}
\end{figure*}

 The training dataset is the same as that in Section 6.
   The test dataset    consists of the researchers  in Set 3, their historical publication quantity  from Set 1,
 and their annual publication quantity  from Set 6.       Its parameters  are
  $t_X=1995$,  and $t_{Y}=2013$.   We only predicted the publication productivity  for the researchers  with no more than $13$ publications at $[T_0,t_X]$, who account  for   98.8\%   of the   researchers in the test  dataset here.
Figs.~\ref{fig10} and \ref{fig11} show
the results of applying the test methods  in Section 6 to the researchers'  productivity predicted by our model.

\end{document}